\documentclass[aps,prl,twocolumn,superscriptaddress]{revtex4-1}

\usepackage{graphicx}  
\usepackage{amsmath} 

\usepackage{graphicx}
\usepackage{amsmath,amssymb,amsfonts}
\usepackage{textcomp}
\usepackage{hyperref}
\usepackage{gensymb}
\usepackage{verbatim}
\usepackage{braket}

\hypersetup{breaklinks=true,colorlinks=true,urlcolor=black}
\usepackage{color}
\usepackage{soul}
\usepackage{gensymb}

\DeclareGraphicsExtensions{.jpg,.pdf,.png,.eps}

\begin{document}


\title{Quantum geometry embedded in unitarity of evolution: revealing its impacts as  geometric oscillation and dephasing in spin resonance and crystal bands.}



\author{B.Q.~Song}
\affiliation{Ames National Laboratory, Iowa State University, Ames, Iowa 50011, USA}
\affiliation{Department of Physics and Astronomy, Iowa State University, Ames, Iowa 50011, USA}
\author{J.D.H.~Smith}
\affiliation{Ames National Laboratory, Iowa State University, Ames, Iowa 50011, USA}
\affiliation{Department of Mathematics, Iowa State University, Ames, Iowa 50011, USA}
\author{T.~Jiang}
\affiliation{Ames National Laboratory, Iowa State University, Ames, Iowa 50011, USA}
\author{Y.X.~Yao}
\affiliation{Ames National Laboratory, Iowa State University, Ames, Iowa 50011, USA}
\affiliation{Department of Physics and Astronomy, Iowa State University, Ames, Iowa 50011, USA}
\author{J.~Wang}
\affiliation{Ames National Laboratory, Iowa State University, Ames, Iowa 50011, USA}
\affiliation{Department of Physics and Astronomy, Iowa State University, Ames, Iowa 50011, USA}


\date{\today}

\begin{abstract}
Quantum Hall effects provide intuitive ways of revealing the topology in crystals, i.e., each quantized ``step" represents a distinct topological state. Here, we seek a counterpart for ``visualizing" quantum geometry, which is a broader concept. We show how geometry emerges in quantum as an intrinsic consequence of unitary evolution, composing a framework compatible with quantum metric and independent of specific details or approximations, suggesting quantum geometry may have widespread applicability. Indeed, we exemplify geometric observables, such as oscillation, dephasing, in magnetic resonance or band driving scenarios. Anomalies, supported by both analytic and numerical solutions, underscore the advantages of adopting a geometric perspective, potentially yielding distinguishable experimental signatures. 
\end{abstract}

\pacs{}


\maketitle

Quantum geometry is a rising field that reveals geometry’s influence (or even dominance) in quantum phenomena \cite{1,2,3,4,5,6,7,8,9,10,11}. The geometry does not refer to the shape of materials nor any tangible apsect, rather to the structure of quantum states \cite{12,13}. 
A hallmark for such abstract formulation is its leading to definition of discrete numbers that characterize material's topological states \cite{14,15}. The topology can be “visualized”, for instance, by Hall conductivities: each step of ${\sigma}_{xy}$ corresponds to a distinct topology. This quantization displays a universal $e^2/h$, independent of material species, thus identifiable in experiments.

Topology is a branch ``quantized geometry". For more general geometry \cite{16,17,18,19,20,21,22,23,24,25,26,27,28,29,30}, we need to uncover other robust observables \cite{29,30,a1,a2} (like counterparts of ${\sigma}_{xy}$ but maybe not quantized). To be identifiable, the phenomenon needs to be ``abnormal", just like the quantized ${\sigma}_{xy}$ going beyond classical understanding.


The goal of this work is two fold. First, show geometry directly embedded in unitarity of evolution, no further precondition. Thus, geometry defined here is a fundamental quantum notion, rather than relying on specific limits \cite{1,8,21}; it gives effect of a geometric origin that is robust and suitable for across-scenario test. Second, find concrete observable to make quantum geometry visible. We underscore distinguishable feature that is qualitatively abnormal to intuitive belief based on perturbation. Note that geometry (analytic result) is more general and accurate than the numerical, and the behavior demonstrated is for proof of principle, rather than specific to models exemplified. 

\textit{The foundation of geometry}. We should see geometry enters for the mere fact that evolution operator is unitary. It endorses the significance of Fubini-Study metric \cite{12,a3,a4}, widely studied as a unitary-invariant quantum distance \cite{7,8,18,a5,a6,a7}; also underscores other unitary invariant geometric quantity, like Haar measure \cite{29,30} (quatnum volume) (Appx. A). Thus, the framework extends the usage of geometry from metric to other aspects, but focus solely on geometry's role within a general and consistent physical ground, not supposed to unify effects yielded under distinct approximations such as adiabaticity \cite{1}, wave-package dynamics \cite{3,21}, thermodynamic limit \cite{8,16,17}, which are irreconcilable.

\begin{figure}
\includegraphics[scale=0.28]{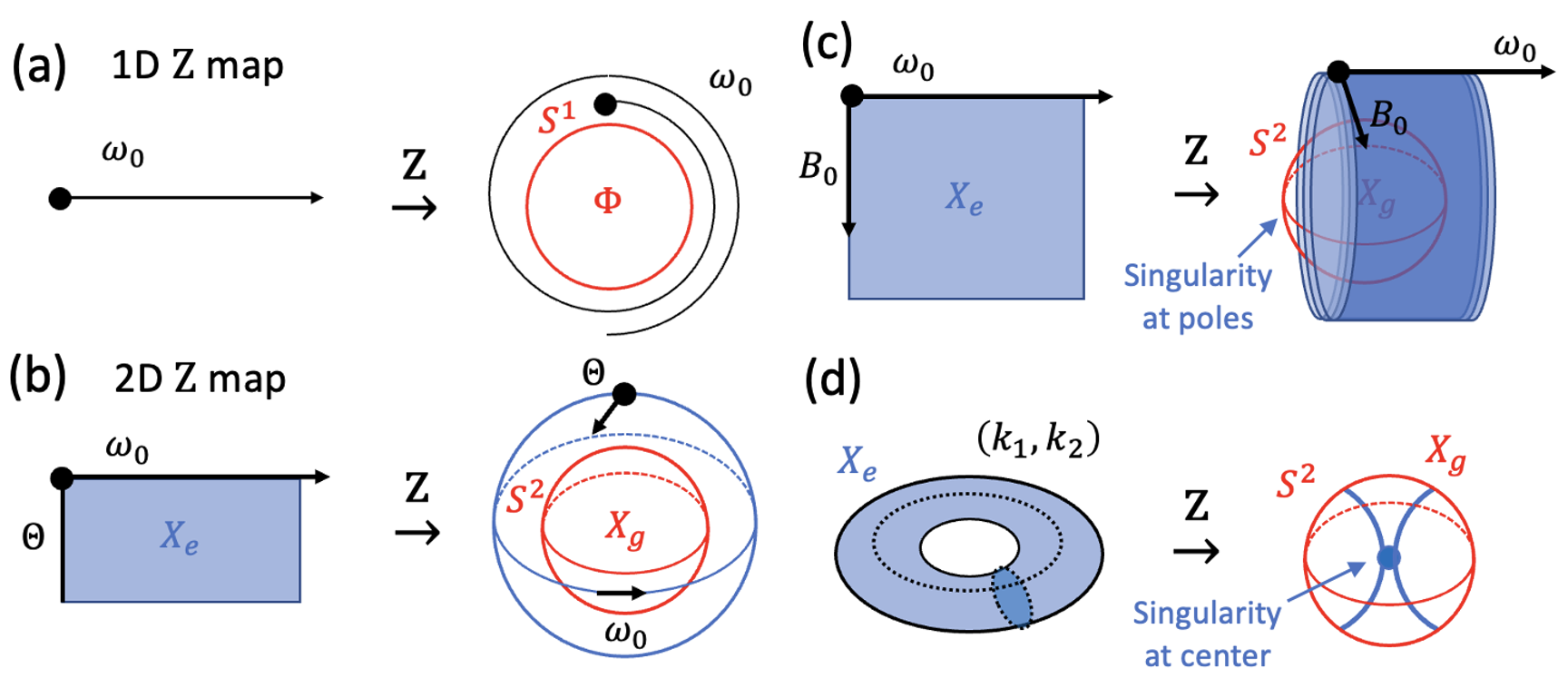}
\caption{\label{fig:epsart}(color online): $Z$-map (restricted to $X_g{\cong}S^2$) under different parameterizations of Hamiltonian. 1D (a) and 2D: (b) isomorphism $S^2{\rightarrow}S^2$; (c) plane $\mathbb{R}^2{\rightarrow}S^2$; (d) Torus $T^2{\rightarrow}S^2$. Singular points are inevitable for (c)(d) for distinct topologies. For (c), imagine a big piece of paper wrapping (winding around) an orange and encountering an annoying point at the pedicle. For (d), surgery has to be performed on the sphere to make it become a torus.\label{f1}}
\end{figure}

Unitary evolution is usually emphasized for its preservation of a state vector's magnitude, i.e., probability, while its geometric meanings as rotations are often overlooked. For example, evolution operator $U$ for spin 1/2 (2D Hilbert space) belongs to SU(2) that involves three Euler angles ${\phi}, {\theta}, {\psi}{\in}[0,2{\pi})$. That means in addition to the dynamic parameters (time $t$ and parameters contained in Hamiltonian $H$), $U$ could be labeled by angles, such as $U({\phi}, {\theta}, {\psi})$. By setting $U(t;B,{\omega},{\cdots})=U({\phi}, {\theta}, {\psi})$, we get smooth maps between the two types of parameters: $(t;B,{\omega},{\cdots}){\mapsto}({\phi}, {\theta}, {\psi})$, called $Z$-\textit{map} denoted as
\begin{equation}
\begin{split}
Z:X_e{\rightarrow}X_g~.
\end{split}
\label{eq1}
\end{equation}
$X_e$ and $X_g$ stand for the spaces of dynamic and geometric parameters. Interestingly, there is intrinsic mismatch between them: $X_e$ often contains boundless parameters, such as $t,~B{\in}(-{\infty},+{\infty})$, while angles in $X_g$ are bounded and periodic. Intuitively, the $Z$-map is about linking two objects of different ``shapes", like a large piece of paper to wrap up an orange. Accurately, a $Z$-map often connects a non-compact space to a compact one, characterized by the dimensions and natures of $X_e$ and $X_g$. Restricting to $\text{dim}X_e=\text{dim}X_g$, we find 1D is simple, just winding around (mod $2{\pi}$, Fig.~\ref{f1}a). In 2D, three topologies of $X_e$ are listed in Fig.~\ref{f1}b-d; in some of them, singularity might be inevitable, such as mapping from cylinder (Fig.~\ref{f1}c) or torus (Fig.~\ref{f1}d) to $S^2$ (a sphere surface), and these singularities may have important indications.

Evidently, quantum evolution is made into numerous categories based on the shapes of $X_e$ and $X_g$ --- geometry and topology have sneaked in for the mere sake of $U$ being unitary. 
The familiar Fubini-Study metric is defined on $S^{2n+1}/U(1)$. If needed, it will enter via $X_g$. For example, $n=1$, corresponds to $S^3/U(1){\cong}S^2$, just the distance on the sphere. The ``$k$-dependence" of metric \cite{a4,a5,a6,a7} corresponds to choosing $X_e$ to be Brillouin zone (BZ), a torus like Fig.~\ref{f1}d, not intrinsic for metric's definition, though. Here, we happen to examine another case Fig.~\ref{f1}c, which gives robust abnormality.

\begin{figure}
\includegraphics[scale=0.42]{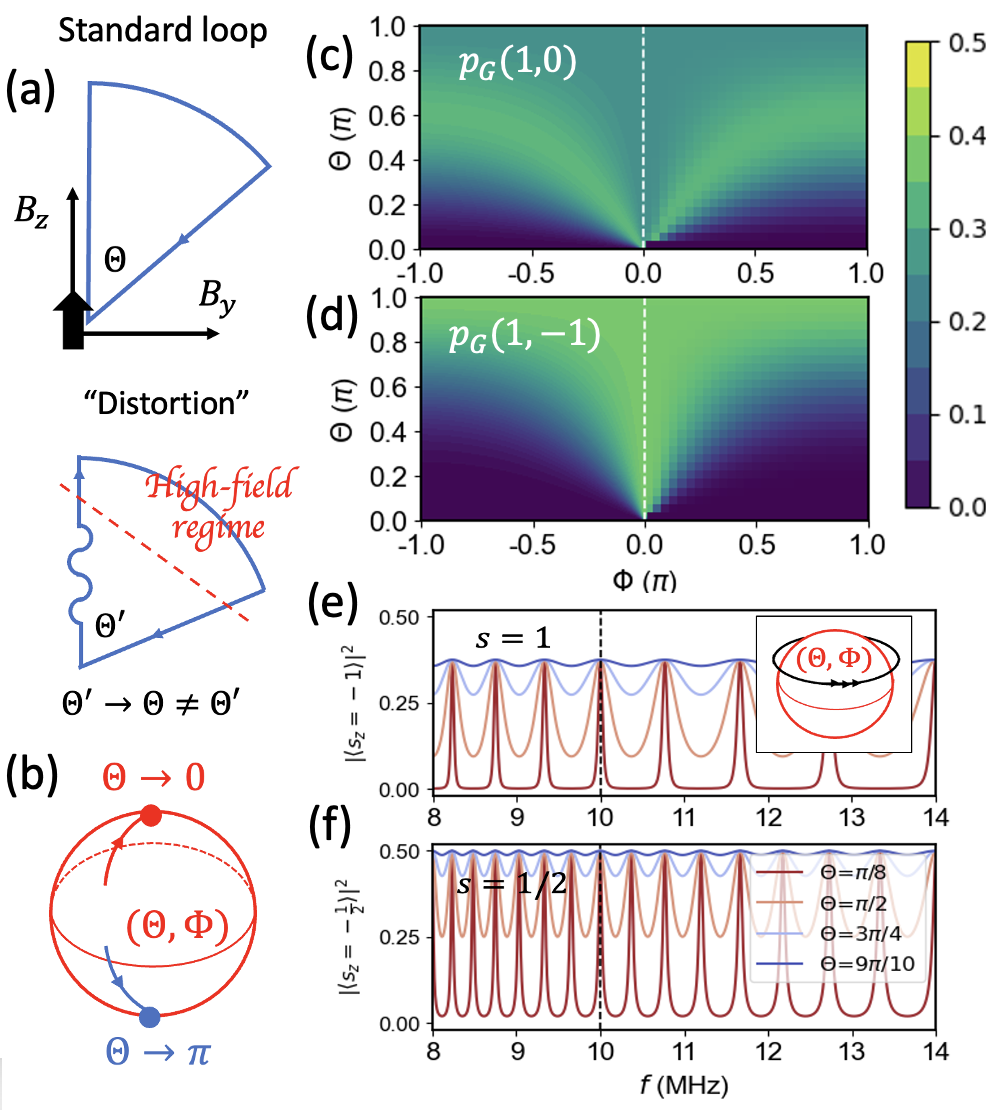}
\caption{\label{fig:epsart}(color online): (a) A standard loop and its distortion due to deviation from the ideal. (b) A quenched pumping ($p_G{\to}0$) is geometrically interpreted as approaching to N-pole of $S^2$, and dephasing is approaching the S-pole. (c)(d) Comparison of analytic (left) and numerical solutions (right, at each pixel $p_G{\approx}p_{n=100}$). (e)(f) Quantum oscillation ($s=1/2,1$) approaches to dephasing tuned by angle ${\Theta}:0{\to}{\pi}$.\label{f2}}
\end{figure}

Consider building a $Z$-map for a spin driven under cyclic ``pie slice" loops of $\textbf{B}(t)$ (Fig.~\ref{f2}a). We recognize $X_e:=({\Theta},{\omega})$, an unbounded ${\omega}:=2{\pi}/{\tau}$ ($\tau$ is the cycle period) together with a bounded $\Theta$ (the vertex angle); other parameters are fixed.

The geometric space is chosen to be $X_g:=({\Theta},{\Phi}){\cong}S^2$, where $\Theta$ is the vertex angle of $\textbf{B}(t)$, while $\Phi$ is a quantum phase arising from evolution. That is,
\begin{equation}
\begin{split}
Z:({\Theta},{\omega}){\mapsto}({\Theta},{\Phi}),
\end{split}
\label{eq2}
\end{equation}
which is the type Fig.~\ref{f1}c. The (one-cycle) evolution operator $\mathcal{U}$ should be expressed with $X_g$. Physically, this corresponds to re-interpretation of dynamic evolution as Hilbert spatial rotation. It shows that (supporting information, SI) spin 1/2 will lead to
\begin{equation}
\begin{split}
\mathcal{U}({\Theta},{\Phi})=\begin{pmatrix} \text{cos}(\frac{\Theta}{2})e^{-i{\Phi}} & -\text{sin}(\frac{\Theta}{2})e^{i{\Phi}} \\ \text{sin}(\frac{\Theta}{2})e^{-i{\Phi}} & \text{cos}(\frac{\Theta}{2})e^{i\Phi} \end{pmatrix}
\end{split}
\label{eq3}
\end{equation}
For spin 1 (a 3D Hilbert space), $\mathcal{U}({\Theta},{\Phi})$ is equal to
\begin{equation}
\begin{split}
\begin{pmatrix} \frac{1}{2}e^{-i{\Phi}}(1+\text{cos}({\Theta}) & -\frac{\sqrt{2}}{2}\text{sin}({\Theta}) & \frac{1}{2}e^{i{\Phi}}(1-\text{cos}({\Theta})) \\ \frac{\sqrt{2}}{2}e^{-i{\Phi}}\text{sin}({\Theta}) & \text{cos}({\Theta}) & -\frac{\sqrt{2}}{2}e^{i{\Phi}}\text{sin}({\Phi}) \\ \frac{1}{2}e^{-i{\Phi}}(1-\text{cos}(\Theta)) & \frac{\sqrt{2}}{2}\text{sin}({\Theta}) & \frac{1}{2}e^{i{\Phi}}(1+\text{cos}({\Theta})) \end{pmatrix}
\end{split}
\label{eq4}
\end{equation}
Notably evolution operators (Eq.~\ref{eq3}, \ref{eq4}) are ``formal" and ``general", because they apply to arbitrary models under a specific Hilbert space. By different models, we plug in different functions $\Theta$, $\Phi$ in terms of original parameters, such as $\omega$, $B$, $t$. (Fortunately, solving explicit forms of these functions can be eluded). Therefore, Eq.~\ref{eq3}, \ref{eq4} attack mutliple models ${\lbrace}H{\rbrace}$ with ``one shot", and the result reflects the common behavior for ${\lbrace}H{\rbrace}$.

Moreover, in view of $Z$-map, such as Eq.~\ref{eq2}, being independent of Hilbert space ($X_e$ and $X_g$ are distinct spaces from Hilbert space), the geometric argument or interpretation will hold for models defined in different dimensions of Hilbert spaces. We demonstrate such ``cross-space" unifying with 2D (Eq.~\ref{eq3}) and 3D (Eq.~\ref{eq4}) (corresponding to 2-fold and 3-fold highest degeneracy). Although they live in different Hilbert spaces, they are described by a common $Z$-map (Eq.~\ref{eq2}). In principle, this applies to arbitrary higher dimensions.




\textit{Quantum oscillation \& dephasing for spins}. To show abnormality, consider a magnetic resonance experiment (Fig.~\ref{f2}a). 
The observable is spin population over eigenstates $|s,s_z{\rangle}$ after $t{\to}{\infty}$ (cycle number $n{\to}{\infty}$). Given an initial state is spin up, i.e., ${\varphi}(t=0)=|\frac{1}{2},\frac{1}{2}{\rangle}$ or $|1,1{\rangle}$. The average population of $n$ cycles is
\begin{equation}
\begin{split}
p_n(s,s_z):=\frac{1}{n}{\sum}_j^n|{\langle}s,s_z|\mathcal{U}^j|{\varphi}(0){\rangle}|^2.
\end{split}
\label{eq5}
\end{equation}
Using Eqs.~\ref{eq3}, \ref{eq4}, we (numerically) evaluate $p_G:=p_{n{\to}{\infty}}{\approx}p_{n=100}$ against $({\Theta},{\Phi})$, presented in the right half of Fig.~\ref{f2}c, d, which suggest $p_G$ peaks at a longitude (the diagram is symmetric with ${\Phi}=0$). Since $({\Theta},{\Phi})$ actually forms $S^2$ surface, if the system traverses along latitudes of $({\Theta},{\Phi})$, pumping should periodically encounter a maximum, forming an oscillation.

The ``motion" on $S^2$ is not a dynamic real-time driving because a point on $S^2$ represents a particular model, and a path is a series of models subject to certain continuous tuning. Particularly, we are interested in frequency $f$ as the tuning parameter, which leads to a pumping oscillation with $f$. Normally, as energy quanta $hf$ increases, pumping will increase before it hits the resonance --- a monotonous trend. By contrast, geometry suggests an oscillation. Within technically plausible frequency of $\textbf{B}(t)$ (Appx. B), spin pumping provides an observable (Fig.~\ref{f2}e, f) to directly ``see" geometry.


The oscillation is abnormal in an energetic viewpoint, because if $hf$ provides more energy, pumping should have been enhanced; besides the probability is fractional (with a fractional ceiling), absent for energetic pumping. Indeed, it originates from geometry. The oscillation could be understood with the hovering around $S^2$ (inset of Fig.~\ref{f2}e), while the fractionality has a deeper geometric origin (shown and proved shortly) and lacks an intuitive picture.

The oscillation exhibits parameter-independent signature, quite identifiable. For example, spin 1/2 is of frequency double with spin 1 under a common condition (Fig.\ref{f2}e, f), which is related to a pure geometric fact that half-integer $s$ gives -1 after rotating $2{\pi}$, while integer $s$ gives +1. Interestingly, this rotation rule for spin can practically be observed by pumping phenomenon at $t{\to}{\infty}$ beyond interference of dynamic phases \cite{31}.

The oscillation can be switched on/off (Fig.~\ref{f2}e, f) by tuning the angle of the $\textbf{B}(t)$ loop (Fig.~\ref{f2}a), thus verifiable in experiments. A flat line (as ${\Theta}{\to}{\pi}$) means pumping probability is insensitive to the quantum phase ${\Phi}$. A basic ingredient in quantum is the phase's influence; now the phase becomes effectless, as though it disappears. The dephasing is shared by both half-integer and integer spins (Fig.~\ref{f2}d, e), and it only depends on ${\Theta}$, thus we call it \textit{geometric dephasing}. 

$Z$-map provides a pathway to analytic solutions for $p_G$ (Sec. 2 of SI). The idea is mapping the evolution path to a special subspace of $X_g$, namely \textit{ergodic subgroup}, \cite{29,30} in which probability density is uniform according to quantum Liouville's theorem recently proved \cite{30}. With integration approaches, we obtain
\begin{equation}
\begin{split}
p_G(1,-1)=\frac{3}{8}\frac{\text{sin}^4(\frac{\Theta}{2})}{(1-\text{cos}^2(\frac{\Theta}{2})\text{cos}^2(\frac{\Phi}{2}))^2}
\end{split}
\label{eq6}
\end{equation}
\begin{equation}
\begin{split}
p_G(1,0)=\frac{\text{sin}^2(\frac{\Theta}{2})}{1-\text{cos}^2(\frac{\Theta}{2})\text{cos}^2(\frac{\Phi}{2})}-\frac{3}{4}\frac{\text{sin}^4(\frac{\Theta}{2})}{(1-\text{cos}^2(\frac{\Theta}{2})\text{cos}^2(\frac{\Phi}{2}))^2}
\end{split}
\label{eq7}
\end{equation}
Analytic and numerical $p_G$ are compared in Fig.~\ref{f2}c, d. The analytic solutions make geometric dephasing a rigorous concept, because all pumping channels shown by Eq.~\ref{eq6}, \ref{eq7} become constant functions of $\Phi$ at ${\Theta}={\pi}$. $Z$-map gives a geometric understanding: approaching to the S-pole of the $S^2$, and since $\Phi$ corresponds to motions along latitudes, it becomes motionless at the S-pole (Fig.~\ref{f2}b), thus causing no effect. Furthermore, it proves fractionality and its exact limit values: $p_G(1,-1){\to}\frac{3}{8}$ and $p_G(1,0){\to}\frac{1}{4}$ at $\Theta{\to}{\pi}$.

Note that it is a limited understanding that geometric effect is just a quantity expressed as a function of quantum metric. Here we see quantum volume is used instead of distance. Moreover, the phenomenon's robustness is related to the inevitable pole, whose math expression is beyond a local analytic function \cite{21,a4,a5,a6,a7}. Indeed, geometry is a broader notion than metric or analytic. 

\textit{Quantum oscillations for bands}. Similar analysis can be made on $H(k)$, given $k$ is conserved. Phonon plays the roles of cyclic fields, and $\textbf{B}=0$ corresponds to gap closing, which might lead to band inversion and thus topological phase transition (TPT). The spin observable is replaced with inter-band pumping, a charge distribution near gap closing $k_0$, formed after a number times of TPT driven by phonons. \cite{32,33,34,35} Consider a two-band model
\begin{equation}
\begin{split}
H(k)=c_H\begin{pmatrix} -{\epsilon}(t)-\text{cos}(k) & -i\text{sin}(k) \\ i\text{sin}(k) & {\epsilon}(t)+\text{cos}(k) \end{pmatrix},
\end{split}
\label{eq8}
\end{equation}
where ${\epsilon}(t)={\epsilon}_0+A_{ph}{\cdot}\text{sin}({\omega}t)$ depicts phonon. 

\begin{figure}
\includegraphics[scale=0.40]{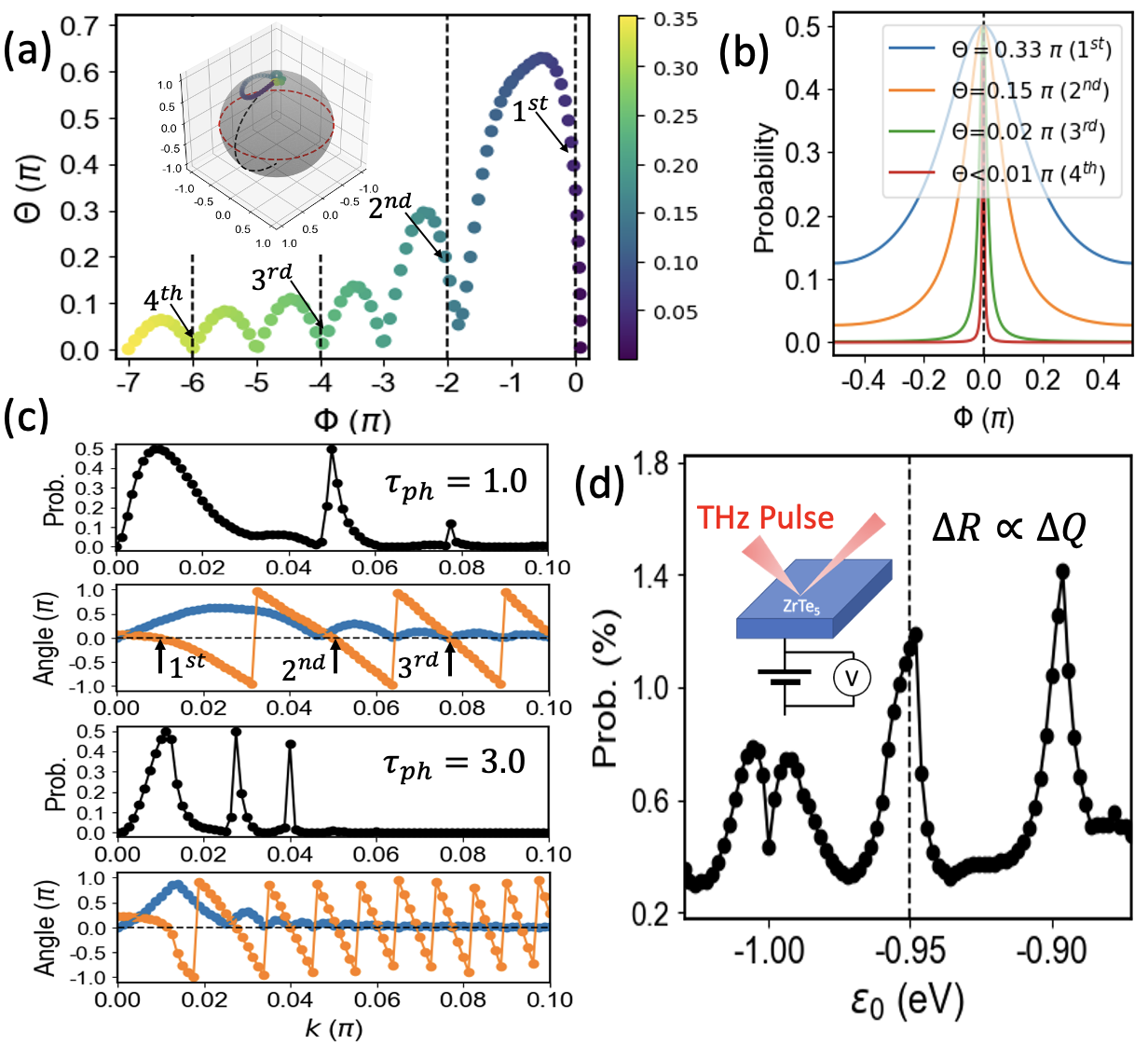}
\caption{\label{fig:epsart}(color online): (a) $k$ serves as the coordinate of a path in $({\Theta},{\Phi})$ space ($S^2$ surface shown by the inset). Every time longitude ${\Phi}=0$ (dashed) is passed, pumping probability $p_G$ gets to maximum (1st, 2nd, 3rd…). The color scales of dots stand for $k$’s value in unit of $\pi$. (b) Analytic solutions of $p_G$ (for spin 1/2). (c) Numerical results of $p_G$ in BZ. ($p_G (k)=p_G(-k))$ subject to different phonon period ${\tau}_{ph}$, and its correspondence to $({\Theta},{\Phi})$ (lower panels). Maximum of $p_G$ occurs at dashed line (lower panels) crossing $\Phi$ (orange dots). Since $\Phi$ is periodic, the “jump” from -1 to 1 is actually continuous. (d) Oscillation of pumping tuned by bias voltages.\label{f3}}
\end{figure}

$Z$-map for bands relies on numerical methods to determine (${\Theta}, {\Phi}$ are now functions of $k$). Given the model Eq.~\ref{eq8}, we employ Trotter decomposition \cite{36,37} to evaluate $U$. Basically, the method discretizes the evolution into steps. Once $U$ is obtained, we use Eq.~\ref{eq3} to deduce $({\Theta},{\Phi})$. The $Z$-map is graphed in Fig.~\ref{f3}a, i.e., $H(k)$ is mapped to $S^2$ surface (inset of Fig.~\ref{f3}a).

Now $k$ plays roles as ${\omega}$ in the spin model, tracing out a path on $S^2$. Analytic solutions (Fig.~\ref{f3}b) suggest the maximum $p_G$ occurs at ${\Phi}=0$. Thus, whenever the longitude ${\Phi}=2n{\pi}$ (dashed lines in Fig.~\ref{f3}a) is passed, pumping peaks. However, as ${\Theta}{\to}0$, the peak width vanishes. Thus, only the first few orders have decent chance to be seen. The simulations are presented in Fig.~\ref{f3}c. In principle, the peaks are equal in heights, but for finite $k$ resolution ($k$-sampling eludes the peak), they appear decreasing --- geometry proves missing features by the numerical simulation.

In analog to spin pumping oscillating with ${\omega}$, charge pumping oscillates with gaps in $k$-space (Fig.~\ref{f3}c): larger gap (minimum gap happens at $\Gamma$) might lead to enhanced probability density of pumping. The abnormality is of the same geometric origin as spin, independent of band or phonon details. We demonstrate the robustness with two phonon periods ${\tau}_{ph}$ (representing different phonon modes). Their patterns are similar except for the longer ${\tau}_{ph}=3$ ps leading to a fast-rolling phase $\Phi$ and thus passing more around $S^2$ than ${\tau}_{ph}=1$ ps (Fig.~\ref{f3}c).

Unlike perturbation, $Z$-map relies solely on unitarity and continuity, providing a distinct picture. However, their opposition could be neutralized from a coarse-grained viewpoint, as the envelop of the oscillations aligns with the tendency suggested by perturbation. Furthermore, geometry circumvents $H$'s detailed forms to yield model-independent behaviors, while energetic analysis like perturbation depends on interaction $\hat{V}$, its strength and truncation orders (or types) \cite{31}, resulting in model-specific outcomes.

To facilitate experimental detection, we make a proposal only dependent on the total probability (sum over BZ). The idea is to employ local site energy ${\epsilon}_0$ (instead of $k$) to drive the oscillation. A plausible tuning for ${\epsilon}_0$ is bias voltage or strain; a suitable system could be ZrTe$_5$ \cite{32,33} a narrow-gapped topological insulator at the vicinity of TPT. The only drawback is averaging might reduce the contrast (Fig.~\ref{f3}d). Fortunately, the pump-probe techniques \cite{25,32,33,34,35} provide sensitive detection of charge pumping by measuring the change of reflectivity ${\Delta}R$ with resolution ${\Delta}R/R$ $10^{-5}{\sim}10^{-6}$.

\textit{Comparison with known effects}. The predicted oscillation should not be confused with an existing term ``quantum oscillation", e.g., the de Haas–Van Alphen (DHVA) effect. The difference lies in (i) it is not is an oscillation in a single system, but an ``up-and-down" displaying in a group of peer systems; (ii) the convolution is not on a ``material" Fermi surface, but on an abstract closed manifold of evolution group, thus not limited to electronic or metallic systems; (iii) the observable is not limited to the response to magnetic fields, but could be multiple. The pumping $p_G$ is one such. It is based on ${\hbar}{\omega}{\ll}\bar{\Delta}$ except for an infinitesimal period of gap closing in the cycle, thus different from the adiabatic limit where ${\hbar}{\omega}{\ll}\bar{\Delta}$ constantly holds, also different from Rabi oscillations \cite{31}, which lead to pumping at ${\hbar}{\omega}{\sim}\bar{\Delta}$.

\textit{Outlook}. $Z$-map proves useful in revealing anomalies and a deep-seated contradiction: the monotonic dependence of pumping on gaps or frequencies, while intuitively satisfying, secretly violates the principles of unitarity and continuity. $Z$-map is independent of energetic parameters and space dimensions, and its strength is to uncover generic behaviors across models. Therefore, a promising direction is by judicious construction of different $Z$-maps (Fig.~\ref{f1}) to discover more such behaviors, analogous to on-going efforts in expressing different quantities in terms of quantum metrics \cite{a4,a5,a6,a7}. However, the focus shifts to general geometry rather than a specific geometric quantity. Another inspiration of $Z$-map is that geometry's influence can go beyond analytic ways, as exemplified by the connection between dephasing and ``poles".

\textit{Conclusion}. First, geometry enters quantum via unitarity, unconstrained by specific approximations or scenarios, leading to generic phenomena and rigorous concepts. 
Second, observables are uncovered for visualizing the influence of quantum geometry in spin/band scenarios. Intuitively speaking, topology is demonstrated by ``steps”, and here geometry can be demonstrated with ``wavy lines”; moreover, these wavy lines are abnormal to conventional belief and can be tuned (for spin, by the loop vertex angle $\Theta$, and for band by band inversion \cite{29,32}), such that their identities could be tested. 




\section{Appendix}
\subsection{A. Generality of $Z$-map \& Its limitation}
(1) \textit{Z-map Unifies different models in different spaces}. The unifying comes in from several tiers. First, different models, say $({\Theta},{\omega}_1), ({\Theta},{\omega}_2),...$ in the original $X_e$, described by the standard ``pie-slice" loops (Fig.~\ref{f2}a) could be mapped to a common point in $X_g$. That is, a point in geometric space $X_g$, say $({\Theta},{\Phi})$, is not just standing for a single model, but an equivalence class of models that lead to a common evolution operator. In other words, the unifying is related to the fact that $Z$-map is non-injective (multiple-to-one). The benefit is that, to examine ``all" models, we not necessarily exhaustively write down all Hamiltonians, but only select one representative in each equivalence class, such as using the ``pie-slice" model. So, instead of covering all possible Hamiltonians, we try to cover all evolution outcomes of Hamiltonians. 
 
Second, following this idea, the unifying could go beyond the standard loop to arbitrary models, as long as they are characterized by a set of parameters. For example, a ``bumpy" loop (Fig.~\ref{f2}a) expands $X_e$ from $(B,{\omega})$ to a higher dimension ${\lbrace}B_j,{\omega}_j{\rbrace}$ (Fig.~\ref{f2} and SI). Notably, increasing $X_e$ will not change $X_g$. The idea is that a model ${\lbrace}B_j,{\omega}_j{\rbrace}$ in the expanded $X_e$ will still find a place on the same $X_g$. Formally, this is expressed with a map ``distortion" $({\Theta},{\Phi}){\mapsto}({\Theta}',{\Phi}')$ (Fig.~\ref{f2}a). That means arbitrary models characterized by a specific $X_e$ could be unified to a common $X_g$.
 
Third, different scenarios, such as spin, band, which are formulated by different $X_e$ spaces, could still be mapped to a common $X_g$. Moreover, the dimension of $X_e$ is independent of the dimension of Hilbert space. In the paper, we show with spin 1/2 and spin 1, which reside in 2D and 3D Hilbert spaces, while they share a common $X_e=(B,{\omega})$. By this example, we see arguments, such as dephasing at ${\Theta}{\mapsto}{\pi}$, can be made across different Hilbert spaces. Otherwise, we would have faced an infinite task of studying different dimensions one by one.

In short, different Hilbert spaces are unified to a common $X_e$; different $X_e$ could be unified under a common $X_g$; then we work on $X_g$ --- This is the ``mechanism" for making generic arguments across scenarios, models, and Hilbert spaces.

(2) \textit{General applicability and phenomena}. Berry phase involves (i) geometric terms (connection, curvature, etc.) (ii) in generic setups. Notably, most protocols for quantum geometry only inherit (i) by exploring further geometric term, such as metric; but lose (ii), because to make metric enter, it relies on specific approximations or scenarios. By contrast, Berry phase is defined in generic parameter space, independent of whether the system is atom or crystal, whether the dispersion is flat or not. 

$Z$-map tries to maintains both (i) and (ii). Firstly, on $X_g$, various geometric quantities (including metric) could be defined; secondly, the existence of $X_g$ merely relies on unitary, thus (ii) is respected too.

Generality is also seen from the general observables uncovered. It is often said quantized ${\sigma}_{xy}$ is general, because it is due to ``continuity", which forbids the wavefunction jumps between topologies. This generality makes the effect robust, insensitive to system’s details. 

Here, $Z$-map predicts robust phenomena, such as geometric oscillation/dephasing, which also arise from ``continuity". Such phenomena might take place in different scenarios and parameter regimes. For example, we show the robustness with different phonon periods ${\tau}_{ph}$ (independent of phonon modes). Besides, the robustness is demonstrated by parameter-independent phenomena, such as geometric oscillation of half-integer $s$ along ${\omega}$ (or $f$) being twice faster than integer $s$ (Fig.~\ref{f2}e, f),

(3) \textit{Relationship with metric-based geometry}. $Z$-map is a comprehensive framework for various geometric quantities to be built in, such as metric (or particularly Fubini-Study metric), measure, curvature. The entrance is just $X_g$ space. Moreover, $X_g$ (solely) arises from unitary evolution, thus naturally underscores the Fubini-Study metric as a unitary invariant. Otherwise, if the formulation merely involves gauge invariant, Fubini-Study metric does not prove irreplaceable, thus not ``physical".

Besides, $Z$-map extends the ways of geometry giving the influence. Most commonly, geometry enters locally, in forms of a local field, such as metric \cite{7,8,9,18,34,35,36,37}, curvature ; or sometimes enters in a global way by an integration of metric \cite{19,24}. The geometric oscillation/dephasing from $Z$-map are both due to global geometry, but not due to an integration of an explicit local field. Besides, it raises the possibility of non-analytic ways of geometry giving its influence, such as the poles when one space tries to cover a different one.

\textit{Limitation of Z-map}. $Z$-map generalizes geometry from metric to measure, and possibly to other aspects. But $Z$-map will not unify every existing scheme as special cases. Because their ``inputs" could be different. $Z$-map is stricter in generality in introducing geometric concepts. In short, they will overlap, but totally inclusive.

Another limitation is that $Z$-map is only to reveal generic behaviors, and detailed information will be ignored. For example, a generic oscillation can be proved by geometry, while the exact locations of peaks or valleys are scenario specific. This is in analog with quantized ${\sigma}_{xy}$, which is due to edge bands protected by topology, thus material-independent; but the detailed dispersion and locations of those edge bands are material-dependent. 

\subsection{B. Viability of experiments}
First, the observable is an oscillation pattern, more detectable than fine features. Second, unlike some fragile phenomena (e.g., spin liquid) living in narrow regimes, the oscillation is robust to model details, easier for material selection and preparation. Third, it is scenario independent, that means it allows crosscheck in different systems from local spins to various narrow-gapped topological insulators, along different axes, such as $\omega$, ${\epsilon}_0$(Fig.~\ref{f2},\ref{f3}). Fourth, the frequency of $\textbf{B}(t)$ is MHz, and the driving of bands is in THz --- both are technically achievable.

In particular, for the band model, we already made preliminary measurement in ZrTe$_5$ \cite{32,33}. The TPT could be realized by $A_{1g}$, $B_{1u}$ phonon modes, which can be excited by THz or optical pulses. The pumping physically is a meta-stable charge configuration after a relatively ``long" time, thus, it is insensitive to the vibration process, as long as the phonon will close up the gap for a number of times. The charge pumping can be measured by transition rate change with ${\sim}10^{-5}$-$10^{-6}$ and sub-ps time resolution. Next, we need to extend the measurement under a biased voltage or applied strains. 

Parameters used for the experimental proposal. In Fig.~\ref{f2}e, f. The frequency $f$ is $1/{\tau}$, where ${\tau}$ is the period of $\textbf{B}(t)$ completing a whole loop. Magnetic resonance experiment is typically performed in a MHz range \cite{38}. Thus, parameters will be chosen to approach that range. By a rough estimate, $\bar{B}=0.01$ T is suitable and technically achievable. Assume ${\Phi}_0=0$ at $f_0=1/{\tau}_0=10$ MHz for a demonstration. (Otherwise, if ${\Phi}=0$ happens at a different frequency, the oscillation simply has an overall shift to the true $f_0$.) Then, ${\Phi}={\Phi}_0+{\Delta}{\Phi}=\bar{E}{\cdot}({\tau}-{\tau}_0)/{\hbar}$ and $\bar{E}={\mu}_B\bar{B}$, where $\bar{E}$ and $\bar{B}$ are the average energy and magnetic field. Some calculations give ${\Phi}=c(1/f-{\tau}_0)$, where $c$ is a coefficent ${\sim}2.8{\pi}{\cdot}10^2$ MHz. Combining $\Phi$ with Eqs.~\ref{eq6},\ref{eq7}, we may plot the oscillations under different $\Theta$. 

As a necessary check, we should remember $f$ cannot be too fast (or equivalently $\bar{B}$ cannot be too small), because the pumping is derived based on $hf{\ll}{\bar{\Delta}}$. In this case, $\bar{B}=0.01$ T leads to $\bar{\Delta}{\sim}10^{-27}$ J and $hf$ is ${\sim}10^{-25}$ J. Thus, the precondition is satisfied. In other words, it is well separated from a Rabi pumping \cite{31}

In Fig.~\ref{f3}, we have adopted $c_H = 0.5$ eV, which gives a Fermi velocity under this model ${\sim}10^5$ m/s, consistent with ZrTe$_5$. More parameters, such as the sampling density in BZ, could be found in SI.



\begin{thebibliography}{99}
\bibitem{1} M. V. Berry, in Geometric Phases in Physics, Advanced Series in Mathematical Physics, Vol. 5, edited by A. Shapere and F. Wilczek (World Scientific, Singapore, 1989), pp. 7–28.
\bibitem{2} R. Resta, The insulating state of matter: A geometrical theory, Eur. Phys. J. B 79, 121 (2011).
\bibitem{3} Q. Niu, M. C. Chang, B. Wu, D. Xiao, Physical effects of geometric phases, World Scientific, Singapore, 2017.
\bibitem{4} D. Vanderbilt, Berry Phases in Electronic Structure Theory: Electric Polarization, Orbital Magnetization and Topological Insulators, 1st Edition, Cambridge University Press, Cambridge, United Kingdom, 2018.
\bibitem{5} N. P. Armitage, E. J. Mele, A. Vishwanath, 
Rev. Mod. Phys. 90, 015001 (2018) .
\bibitem{6} J. Orenstein, J. E. Moore, T. Morimoto, D. H. Torchinsky, J.W. Harter, and D. Hsieh, 
Annu. Rev. Condens. Matter Phys. 12, 247 (2021).
\bibitem{7} J. Ahn, G.-Y. Guo, N. Nagaosa, A. Vishwanath, Nature Phys. 18, 290-295 (2022).
\bibitem{8} K. E. Huhtinen, J. Herzog-Arbeitman, A. Chew, B. A. Bernevig, P. T\"orm\"a, 
Phys. Rev. B 106, 014518 (2022).
\bibitem{9} P. T\"orm\"a, 
Phys. Rev. Lett. 131, 240001 (2023).
\bibitem{10} J. G. Checkelsky, B. A. Bernevig, P. Coleman, Q. Si, S. Paschen, Nat. Rev. Mater. https://doi.org/10.1038/s41578-023-00644-z (2024) 
\bibitem{11} B. Q. Song, J. D. H. Smith, J Wang, Position operators in terms of converging finite-dimensional matrices: Exploring their interplay with geometry, transport, and gauge theory, arXiv:2403.02519, (2024).

\bibitem{12} J. P. Provost, G. Vallee, 
Commun. Math. Phys. 76, 289 (1980).
\bibitem{13} J. Anandan, Y. Aharonov, 
Phys. Rev. Lett. 65, 1697 (1990).

\bibitem{14} Y. Hatsugai, Phys. Rev. Lett. 71, 3697 (1993).
\bibitem{15} X.-L. Qi, T. L. Hughes, S.-C. Zhang, Phys. Rev. B 78, 195424 (2008).

\bibitem{16} A. Gao et al., 
Science 381, 181 (2023).
\bibitem{17} J. Herzog-Arbeitman et al., Topological heavy Fermion principle for flat (narrow) bands with concentrated quantum geometry, arXiv 2404.07253, 2024
\bibitem{18} J. Ahn, N. Nagaosa, 
Phys. Rev. B 104, L100501 (2021).
\bibitem{19} S. Peotta, P. T\"orm\"a, 
Nat. Commun. 6, 8944 (2015).
\bibitem{20} J. Herzog-Arbeitman, V. Peri, F. Schindler, S. D. Huber, B. A. Bernevig, 
Phys. Rev. Lett. 128, 087002 (2022).
\bibitem{21} M. C. Chang, Q. Niu, Phys. Rev. B 53, 7010 (1996).
\bibitem{22} A. Jaoui, I. Das, G., Di Battista, J. Diez-Merida, X. Lu, K. Watanabe, T. Taniguchi, H. Ishizuka, L. Levitov, D. K. Efetov, 
Nat. Phys. 18, 633–638 (2022).
\bibitem{23} L. Balents, C. R. Dean, D. K. Efetov, A. F. Young, 
Nat. Phys. 16, 725-733 (2020)
\bibitem{24} H. Tian, X. Gao, Y. Zhang, S. Che, T. Xu, P. Cheung, K. Watanabe, T. Taniguchi, M. Randeria, F. Zhang, C. N. Lau, M. W. Bockrath, 
Nature 614, 440 (2023).
\bibitem{25} B. Cheng et al., Chirality manipulation of ultrafast phase switches in a correlated CDW-Weyl semimetal, Nat. Commun. 15, 785 (2024).
\bibitem{26} G. R. Stewart, 
Rev. Mod. Phys. 56, 755 (1984).
\bibitem{27} Y. Zeng et al., 
Nature 622 69 (2023).
\bibitem{28} E. Andrei, D. Efetov, P. Jarillo-Herrero, A. MacDonald, K. Mak, T. Senthil, E. Tutuc, A. Yazdani, A. Young, 
Nat. Rev. Mater. 6, 201–206 (2021).
\bibitem{29} B. Q. Song, J. D. H. Smith, L. Luo, J. Wang, Phys. Rev. B 105, 035101 (2022).
\bibitem{30} B. Q. Song, J. D. H. Smith, L. Luo, J. Wang, Phys. Rev. B 109, 144301 (2024).

\bibitem{a1} H. Tian, X. Gao, Y. Zhang, S. Che, T. Xu, P. Cheung, K. Watanabe, T. Taniguchi, M. Randeria, F. Zhang, C. N. Lau, and M. W. Bockrath, 
Nature 614, 440 (2023).
\bibitem{a2} G. Sala, M. T. Mercaldo, K. Domi, S. Gariglio, M. Cuoco, C. Ortix, and A. D. Caviglia, The quantum metric of electrons with spin-momentum locking, 
arXiv:2407.06659 (2024).
\bibitem{a3} M. Kolodrubetz, D. Sels, P. Mehta, A. Polkovnikov, 
Physics Reports 697, 1 (2017).
\bibitem{a4} J. Yu, B. A. Bernevig, Raquel Queiroz, Enrico Rossi, P\"aivi T\"orm\"a, B.-J. Yang, arXiv: 2501.00098 (2024).
\bibitem{a5} S. A. Chen, K. T. Law, 
Phys. Rev. Lett. 132, 026002 (2024).
\bibitem{a6} A. Avdoshkin, J. Mitscherling, J. E. Moore, The multi-state geometry of shift current and polarization, arXiv:2409.16358 (2024).
\bibitem{a7} P. Virtanen, R. P. S. Penttil\"a, P. T\"orm\"a A. D\'iez-Carl\'on, D. K. Efetov, T. T. Heikkil\"a, Superconducting junctions with flat bands, arXiv:2410.23121 (2024).

\bibitem{31} D. Griffiths, Introduction to Quantum Mechanics (2nd ed. Cambridge university press 2005)


\bibitem{32} C. Vaswani, L.-L. Wang, D. H. Mudiyanselage, Q. Li, P. M. Lozano, G. D. Gu, D. Cheng, B. Song, L. Luo, R. H. J. Kim, C. Huang, Z. Liu, M. Mootz, I. E. Perakis, Y. Yao, K. M. Ho, and J. Wang, Phys. Rev. X 10, 021013 (2020).
\bibitem{33} L. Luo, D. Cheng, B. Song, L.-L. Wang, C. Vaswani, P. M. Lozano, G. Gu, C. Huang, R. H. J. Kim, Z. Liu, J.-M. Park, Y. Yao, K. Ho, I. E. Perakis, Q. Li, and J. Wang, Nat. Mater. 20, 329 (2021).
\bibitem{34} B. Q. Song, X. Yang, C. Sundahl, J.-H. Kang, M. Mootz, Y. Yao, I. E. Perakis, L. Luo, C. B. Eom, J. Wang, Ultrafast martensitic phase transition driven by intense terahertz pulses, 3, 0007 (2023)
\bibitem{35} X. Yang, X. Zhao, C. Vaswani, C. Sundahl, B. Song, Y. Yao, D. Cheng, Z. Liu, P. P. Orth, M. Mootz, J. H. Kang, I. E. Perakis, C.-Z. Wang, K.-M. Ho, C. B. Eom, J. Wang, Phys. Rev. B 99, 094504 (2019)
\bibitem{36} T. Jiang et al., 
Communications Physics 6, 297 (2023)
\bibitem{37} H. F. Trotter, 
Proc. Am. Math. Soc. 10, 545 (1959).
\bibitem{38} M. H. Levitt, Spin Dynamics: Basics of Nuclear Magnetic Resonance, (2nd Ed. Wiley 2008)

\end{thebibliography}
\end{document}